\documentstyle[twocolumn,aps,epsfig]{revtex}
\def\ltsim{\vbox {\hbox{\lower .8\baselineskip \hbox{$<$}} \break
                 \hbox{\lower 0.2\baselineskip \hbox{$\sim$}} } }

\begin{document}
\draft

\twocolumn[\hsize\textwidth\columnwidth\hsize\csname 
@twocolumnfalse\endcsname
                                 
\title{Heat conduction and Wiedemann-Franz Law in disordered Luttinger 
Liquids}

\author{Mei-Rong Li$^1$ and E. Orignac$^2$}
\address{$^1$Department of Physics, University of Guelph, Guelph, 
Ontario, Canada N1G 2W1\\
$^2$LPTENS, CNRS UMR8549, 24, Rue Lhomond - 75231 Paris Cedex 05, 
France} 
       
\maketitle
\parindent 2em

\begin{abstract}
We consider heat transport in a Luttinger liquid (LL) with weak 
disorder and study the Lorenz number for this system. We start at 
a high-$T$ regime, and calculate both the electrical and thermal 
conductivities using a memory function approach. The resulting Lorenz 
number $L$ is independent of $T$ but depends explicitly on the LL 
exponents. Lowering $T$, however, allows for a renormalization of  
the LL exponents from their bare values by disorder, causing a 
violation of the Wiedemann-Franz law. Finally, we extend the 
discussion to quantum wire systems and study the wire size 
dependence of the Lorenz number. 
\end{abstract}
]

The Wiedemann-Franz (WF) law states that for electrons {\it elastically} 
scattered by impurities the ratio of the thermal ($\kappa$) and electrical 
($\sigma_c$) conductivities normalized by $T$ is a {\it universal} number 
$L_0=\kappa/\sigma_c T = \pi^2/3e^2$ called the Lorenz number. This law 
is obeyed for electrons in conventional metals \cite{ziman_phonon_book}, 
and inclusion of electronic interactions that lead to the standard 
Landau Fermi-liquid (FL) behavior does not modify $L_0$ 
\cite{castellani_lorenz}, as their effect on $\kappa$ and $\sigma_c$ 
scales away when the particle-particle scattering rate is much smaller 
than the impurity scattering rate.

A fundamental question arises whether the WF law survives in strongly 
correlated electron systems with non-FL ground states. Interest in this 
question has been  recently revived by an observation of violation of WF 
law in high-$T_c$ cuprates \cite{hill} which  are widely believed to be  
non-FL systems. An important and tractable example of a non-FL is a
one-dimensional (1d) Luttinger liquid (LL) with separate charge ($\rho$) 
and spin ($\sigma$) bosonic collective excitations 
\cite{emery_revue_1d,solyom_revue_1d,vondelft_revue1d} instead of 
fermionic quasiparticles. Heat being carried by both spin and charge 
excitations, there is a potential for violation of WF law in a LL. 
However, a potential complication in the LL case is that even a weak 
disorder can cause Anderson localization 
\cite{apel_spinless,apel_loc,apel_impurity_1d,scha,giamarchi_loc_ref}.

Experimentally, thermal transport in (quasi-)1d has been investigated 
in spin-chain, \cite{spinchains} spin-Peierls, \cite{spinpeierls} and 
spin-ladder materials \cite{ladder}. There are also some early 
measurements of  heat conduction in quasi-1d organic conductors
\cite{choi_thermal_tmtsf}. Theoretically, although $\sigma_c(T)$ in the 
LLs has been extensively discussed 
\cite{luther_conductivite_disorder,giamarchi_umklapp_1d,rosch_transport1d} 
with memory function methods \cite{goetze}, the thermal conductivity and 
the WF law have only been addressed in two contexts. First, Kane and Fisher 
(KF) \cite{kane_thermal_1d} examined spinless LLs with a single strong 
impurity for repulsive electronic interactions, and found a 
$T$-independent but nonuniversal Lorenz number $L^{({\rm KF})}$. The 
generalization of this problem to the bulk disorder case is however not 
straightforward \cite{giamarchi_impurities} (as discussed below). Second, 
Fazio {\it et al.} \cite{fazio_thermal_1d} and Krive \cite{krive_thermal1d} 
have studied thermal transport in a mesoscopic 1d wire connected to leads. 
They showed that even for a clean wire the contact with leads causes
violation of the WF law except at sufficiently low $T$. However, the 
effect of backward scattering by finite-density impurities has not been 
considered in the latter case. 

In this letter, we first examine the WF law for an infinite LL with a
nonzero concentration of impurities. Unlike KF, we calculate $\kappa(T)$ 
and $\sigma_c(T)$ in the high-$T$ regime where impurity scattering can be 
treated by the memory function method
\cite{goetze,luther_conductivite_disorder}. The resulting Lorenz number 
is independent of $T$ but depends on the LL exponents. It differs from the 
one obtained by KF. As $T$ goes down, we show that the Lorenz number 
acquires $T$ dependence as the consequence of instability towards 
Anderson localized state. Such a violation of the WF law is in sharp
contrast to the higher-d FL situation. Finally we examine the effects  
of contact in a realistic quantum wire measurement and discuss the 
wire-size dependence of the Lorenz number.

{\it Conductivities from Memory function method.} 
A LL of spin-1/2 fermions in the presence of impurities is described 
by the following Hamiltonian \cite{giamarchi_loc_ref}
\begin{eqnarray}
&& H=H_{LL}+H_{\rm imp}=\int dx \, {\cal H}(x), \label{ham} \\
&& H_{\rm LL}= \sum_{i=\rho,\sigma} \, \int {dx\over {2\pi}} 
\big\{u_iK_i [\pi \Pi_i(x)]^2 + {u_i\over K_{i}} 
\big[ \partial_x \phi_i(x)\big]^2\big\},    \label{hll}   \\
&& H_{\rm imp} = -\frac{\sqrt{2}}{\pi}\int dx\, \eta(x) \partial_x 
\phi_\rho         \nonumber \\ 
&& \;\;\;\;  + \int dx \bigg\{\frac{\xi(x)}{\pi a} e^{i \sqrt{2}  
\phi_\rho(x) }  \cos[\sqrt{2} \phi_\sigma(x)] + \text{ H.c.} \bigg\},   
\label{himp}
\end{eqnarray}
where $K_i$ are the LL exponents, $u_i$ the velocities of the 
excitations, $a$ the lattice constant, and $\Pi_i(x)$ and $\phi_i(x)$ 
are canonical momenta and coordinates, respectively. $\xi(x)$ in Eq. 
(\ref{himp}) is the $2k_F$ component of the impurity potential,  
assumed to have a Gaussian distribution with zero average,
$\overline{\xi(x)\xi^*(x')}=D_\xi \, \delta(x-x').$ $\eta(x)$ is the 
forward scattering component. The charge and heat currents 
$J_c=\int dx {\cal J}_c(x)$ and $J_Q=\int dx {\cal J}_Q(x)$ are 
obtained from the continuity equations $\partial_x {\cal J}_c(x)
+\partial_t n(x)=0$ and $\partial_x {\cal J}_Q(x)+\partial_t 
{\cal H}(x)=0$, where $n(x)=-\frac{\sqrt{2}} \pi \partial_x 
\phi_\rho(x)$ is the electron density operator. We find
\begin{eqnarray}
{\cal J}_c(x) & =& \sqrt{2} \, u_\rho K_{\rho} \Pi_\rho(x), 
\label{chargecurrent}\\
{\cal J}_Q(x) &=& -u_\rho^2 \Pi_\rho \partial_x\phi_\rho 
-u^2_\sigma \Pi_\sigma \partial_x\phi_\sigma + {\cal J}^\rho_\eta(x),
\label{heatcurrent}
\end{eqnarray}
where ${\cal J}^\rho_\eta(x)=\sqrt{2} u_\rho K_\rho \eta(x)\Pi_\rho(x).$
We note that the transformation $\tilde{\phi_\rho}(x)=\phi_\rho(x) -
\sqrt{2}{K_\rho\over u_\rho} \int^{x}\eta(x') dx'$ eliminates $\eta$ from 
both $H_{\text{imp}}$ and $J_Q$, implying a null effect of $\eta(x)$ on 
$\sigma_c$ and $\kappa$. Thus we set $\eta(x) \equiv 0$ hearafter. 
Eqs. (\ref{chargecurrent}) and (\ref{heatcurrent}) also show that the 
heat and charge current operators are quadratic and linear in boson 
operators, respectively. This is crucial in resulting in deviation of the 
Lorenz number from the universal number at high $T$ even for spinless 
fermion case as shown below.

In the absence of impurities, both $J_c$ and $J_Q$ are conserved
currents. In fact, $J_Q$ is (up to a prefactor) the total momentum
of the system, and its conservation results from translational
invariance. With impurities, neither $J_c$ and $J_Q$ are conserved and
finite conductivities can be expected\cite{rosch_transport1d}. 
At sufficiently high $T$, quantum effects are cut off by inelastic 
thermal processes, and disorder scattering can be treated within the 
Born approximation. This yields conductivities to leading order in 
$D_\xi^{-1}$. A convenient formalism to do this is the memory function 
method \cite{goetze,giamarchi_umklapp_1d,rosch_transport1d}, in which 
the finite-frequency electrical and thermal conductivities are expressed
in terms of the memory functions $M_j(\omega)=\chi_j^{-1} \, \omega^{-1} \,
\big[ \langle\langle F_j;F_j\rangle\rangle_\omega -\langle\langle 
F_j;F_j\rangle\rangle_{\omega=0} \big]$ $(j=c,Q)$ as
\begin{eqnarray}
&&\sigma_c(\omega) = i\,e^2 \chi_c [\omega+ M_c(\omega)]^{-1}, 
\label{sigma} \\
&&\kappa(\omega) = i\, \chi_Q \, T^{-1} 
[\omega+ M_Q(\omega)]^{-1},
\label{kappa} 
\end{eqnarray}
with $\chi_j  =  T \int^{1/T}_0 d\lambda \langle J_j(0)^\dagger
J_j(i\lambda) \rangle $ the current static susceptibilities, 
$F_j(x)=[{\cal J}_j(x),H]$, and $\langle\langle F_j;F_j  
\rangle\rangle_{\omega} = - \int dx dx' \int^\infty_0 dt 
\,e^{i\omega t}\; \overline{\langle [F_j(x,t), F_j(x',0)] \rangle_H}.$
To leading order in $D_\xi$,  both $\chi_c$ and $\chi_Q$ take their
pure LL values, $\chi_c \simeq {2 \over \pi} \,u_\rho K_\rho, $ 
$\chi_Q \simeq {\pi\over 3} (u_\rho+u_\sigma) \, T^2$, and 
\begin{eqnarray}
&& \langle\langle F_j;F_j  \rangle\rangle_{\omega}
\simeq D_\xi \, 2^{K_t+2} (\pi a)^{K_t-2}  u_\rho^{-K_\rho} \, 
u_\sigma^{-K_\sigma}\, A_j(\omega)  \nonumber  \\
&& \;\;\;\;\;\;\; \times  \sin \left({\pi K_t\over 2}\right)
B\left( {K_t\over 2}-{i\omega\over 2\pi T}, 1-K_t\right)\, 
T^{-\alpha_j},    \label{ff}
\end{eqnarray}
where 
$K_t=K_\rho+K_\sigma,$ $B(x,y)$ is the Euler Beta function, 
$\alpha_c=1-K_t$, $\alpha_Q=\alpha_c-2$, $ A_c(\omega)= K^2_{\rho} 
u_\rho^{2}/\pi,$ and $A_Q(\omega) = \pi \big[(K^2_t/4 +
\big(\omega/2\pi T\big)^2 \big] (K_\rho u^2_\rho + K_\sigma 
u^2_\sigma)/[K_t(K_t+1)].$
Inserting Eqs. (\ref{ff}) into Eqs. (\ref{sigma}-\ref{kappa}), and after 
some straightforward algebra, we find that the dc conductivities become
\begin{eqnarray}   \label{eq:sigma1-1}
\sigma_c(T) &\simeq & \frac{2e^2\Gamma(K_t)}{\pi\, \Gamma^2(K_t/2)}
{a \over \pi {\cal D}} \left(\frac{2\pi a T}{u_\sigma}\right)^{2-K_t}
= e^2 \,l_{\text{el}}(T), \\
\label{eq:kappa1-1}
\kappa(T)&\simeq & \frac{2\pi^2 T}{9} \frac{1+K_t}{K_t}
\frac{(u_\rho+u_\sigma)^2} {u_\rho^2 K_\rho + u_\sigma^2 K_\sigma} 
 \frac{\Gamma(K_t)}{\Gamma^2(K_t/2}    \nonumber \\
&& \times \frac{a}{\pi {\cal D}} \left(\frac{2\pi a T}
{u_\sigma}\right)^{2-K_t}\propto  C_v(T) \, l_{\text{el}}(T),
\end{eqnarray}
where ${\cal D}=(2D_\xi a/\pi u_\sigma^2) (u_\sigma/u_\rho)^{K_\rho}$ 
is a dimensionless disorder parameter \cite{giamarchi_loc_ref},
$\Gamma(x)$ the Gamma function, $C_v(T)\propto T$ the specific heat 
of a pure LL, and $l_{\text{el}}(T)$ the elastic mean free path 
\cite{apel_loc}. Both $\sigma_c$ and $\kappa$ exhibit power-law behavior 
in $T$ with nonuniversal exponents.

{\it Lorenz number at high $T$.} 
From Eqs. (\ref{eq:sigma1-1}) and (\ref{eq:kappa1-1}), one has 
\begin{eqnarray}  \label{eq:lorenz}
L&=&{\kappa\over \sigma_c T} = { \pi^2\over 9e^2}\, 
{(u_\rho+u_\sigma)^2 \over u_\rho^2 K_{\rho} + u_\sigma^2 K_\sigma } 
\,  {1+K_t\over K_t}. 
\end{eqnarray}
$K_\rho$, $K_\sigma$, $u_\rho$ and $u_\sigma$ are bare intrinsic 
parameters, so $L$ is independent of $T$, and a generalized WF law is 
obeyed. In the noninteracting case, $K_\rho=K_\sigma=1$, 
$u_\rho=u_\sigma=v_F$, the universal Lorenz number $L=L_0$ is recovered 
from Eq. (\ref{eq:lorenz}). 

The result (\ref{eq:lorenz}) can be reduced to the {\it spinless} 
LL case, by making $u_\rho=u_\sigma=u$, $K_\rho=K_\sigma=K$. We obtain
\begin{eqnarray}  \label{eq:lorenz_spinless}
L'= (\pi^2/ 9e^2) \, (K^{-2}+ 2 K^{-1}). 
\end{eqnarray}
It follows that $L' <L_0$ for an attractive interaction case $K>1$, 
as a result of tendency towards a superconducting state with high 
electrical conduction but poor thermal conduction; While for a repulsion 
case $K<1$, $L'>L_0$, indicating tendency towards a weakly pinned charge 
density wave (CDW) state with better heat transport than charge transport. 
It is instructive to compare $L'$ with $L^{\rm (KF)}=(\pi^2/e^2) 
(K^2 +2K)^{-1}$ obtained by KF for a single strong impurity at $K<1$ 
\cite{kane_thermal_1d}. We see that $L'-L^{({\rm KF})}=2(K-1)^2/9K^2(K+2) 
\geq 0$. From physical point of view, $L^{(KF)}$ and $L'$ can be roughly 
understood as the results for a strongly pinned and a weakly pinned CDW 
phase, respectively, so that we should expect a {\it larger} $L^{(KF)}$ 
than $L'$. This puzzle can be resolved by noting that some Hamiltonian 
terms causing heat but no charge conduction are neglected in the 
derivation of the tunneling Hamiltonian used to calculate $L^{(KF)}$ 
\cite{kane_thermal_1d}. Including these terms explicitly for $K=1/2$ 
indeed leads to  \cite{kane_thermal_1d} $\tilde{L}^{(KF)}={3\over 2}
L^{(KF)}>L'$.

We would also like to remark that the single strong impurity case 
considered by KF corresponds to a low energy fixed point which cannot 
be reached from the limit of weak impurity scatterers at high 
concentration $c$ we consider\cite{giamarchi_impurities}. The KF fixed 
point is realized when the renormalized impurity strength 
$t_B(l)=e^{(2-2K)l} t_B(0)$  becomes of the order of the high energy 
cutoff $W\sim u/a$. This requires the renormalized length 
$ae^{l'}=a[W/t_B(0)]^{1/(2-2K)}$ to remain still much smaller than the 
inter-impurity distance $1/c$. In the high impurity concentration limit, 
$c\to \infty$, $t_B \to 0$ with ${\cal D}=ct_B^2$ fixed, obviously such 
regime cannot be observed.
 
{\it Lorenz number with decreasing $T$.}
The enhanced quantum interference effects, which are responsible for 
occurrence of the Anderson localization at $T<T_{\rm loc}=
u_\sigma/l_{\rm loc}$ ($l_{\rm loc}=a{\cal D}^{-1/(3-K_t)}$ 
\cite{giamarchi_loc_ref}), lead to renormalization of $K_\rho$, 
$K_\sigma$ and ${\cal D}$. Now we study the influence of such 
renormalization on $L$ for $T$ still much larger than $T_{\rm loc}$,
so that a perturbative renormalization group (RG) theory can be used.
We first neglect renormalization of $K_\rho$ and $K_\sigma$ by ${\cal D}$. 
The RG flow equation determining the renormalization of ${\cal D}$ reads 
\cite{giamarchi_loc_ref} $d {\cal D}(l)/dl=(3-K_t) {\cal D}(l),$ which 
leads to 
\begin{equation}  \label{eq:disorder}
{\cal D}(l)/{\cal D}=e^{(3-K_t)l}. 
\end{equation}
It is clear that $K_t=3$ defines a metal-insulator transition (MIT) 
line: for $K_t<3$, ${\cal D}(l)$ grows exponentially with renormalized 
length, and the Anderson localization will take place eventually; 
Whereas for $K_t>3$, ${\cal D}(l)$ flows to zero, corresponding to 
a delocalized phase. By defining a $T$-dependent scale $l^*(T)$ from 
$ae^{l^*(T)}=u_\sigma/(2\pi T)=l_{\rm th}(T)$ ($l_{\rm th}$ the thermal 
length), we see that Eq. (\ref{eq:disorder}) yields 
$(a/\pi {\cal D}) (2\pi a T/u_\sigma)^{2-K_t} = a e^{l^*(T)}/[ 
\pi {\cal D}(l^*)],$ which allows for a simple RG interpretation for 
$\sigma_c$ and $\kappa$ in Eqs. (\ref{eq:sigma1-1}-\ref{eq:kappa1-1}): 
They can be obtained by applying the RG flow up to the scale $l^*(T)$ at 
which $T$ becomes order of the energy cutoff $\pi u_\sigma/a$. At this 
scale, thermal effects suppress quantum effects, and the use of the 
memory function method is justified. One can then perform a memory 
function calculation in which the bare parameters ${\cal D}$ and $a$ 
are replaced by ${\cal D}(l^*)$ and $ae^{l^*}$, respectively. 

A similar analysis can be made for inclusion of renormalization of 
$K_\rho$ and $K_\sigma$ by ${\cal D}$. We start from the bare parameters 
$K_\rho(0)$, $K_\sigma(0)$ and $D_\xi(0)$ and follow their RG flow, being 
described by Eqs. (3.4) in Ref. \onlinecite{giamarchi_loc_ref}, up to the 
scale $l^*(T)$. Since $D_\xi(l^*)\ll 1$ still holds for 
$T\gg T_{\rm loc}$, but the quantum phase coherence has started to lose, 
Eqs. (\ref{eq:sigma1-1}-\ref{eq:kappa1-1}) are valid again \cite{footnote2}, 
but with the bare $K_\rho$ and $K_\sigma$ replaced by the renormalized 
ones, $K_\rho(l^*)$ and $K_\sigma(l^*)$, respectively. The resulting 
RG-improved $L$ acquires $T$ dependence through $K_\rho$ and $K_\sigma$, 
and the generalized WF law holding at high $T$ breaks down. This is in 
contrast to a higher-d FL case, in which such a renormalization is always 
negligible \cite{castellani_lorenz}.

\begin{figure}[h]
\centerline{\epsfig{file=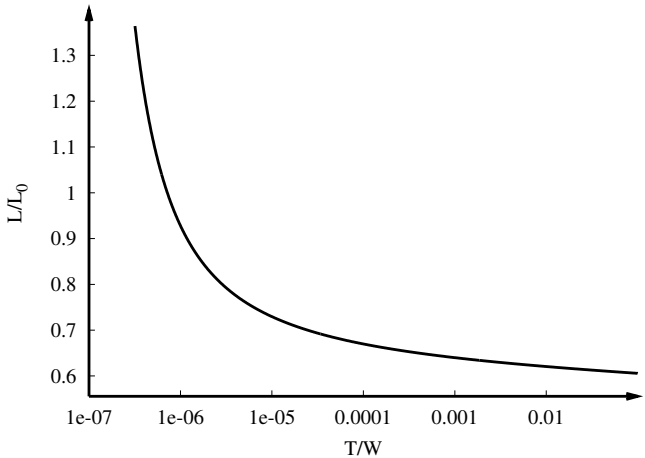,angle=0,width=9cm}}
\caption{Lorenz number as a function of temperature 
close to the MIT line for spinless fermions. $W\sim u/a$ is a high 
energy cutoff. The initial conditions are $ K(0)=3/2, 
{\cal D}(0)=10^{-2}$.}
\label{fig:lorenz}
\end{figure}

We proceed with a discussion of the spinless fermion case for simplicity. 
The above analysis suggests that $L'$ in Eq. (\ref{eq:lorenz_spinless}) 
as a function of the initial parameter $K$ becomes  
\begin{eqnarray} \label{eq:lorenz_T}
L(T)=(\pi^2/ 9e^2)\, \{[K(l^*)]^{-2}+ 2[K(l^*)]^{-1}\}, 
\end{eqnarray}
where $K(l^*)$ can be obtained from the RG flow equation 
\cite{giamarchi_loc_ref}, $dK(l)/dl=-K^2(l) {\cal D}(l)/2$. For a generic 
initial $K$ away from the MIT line, the RG flow line in the $K$-${\cal D}$ 
plane is almost vertical, and the deviation of $L$ from its high-$T$ value 
is in order of ${\cal D}$ which is insignificant unless $T$ becomes close 
to $T_{\text{loc}}$. Alternatively, considering the second order terms in 
${\cal D}$ in $\sigma_c$ and $\kappa$ one also obtains a $T$-dependent  
power-law correction term to $L(T)$ with the exponent predicted by the 
RG method. 

However, in the vicinity of the MIT line, the RG flow is no longer 
vertical, leading to a faster variation of $L$ with $T$. $|K(l^*)-K|$ is 
found to be proportional to $\sim \sqrt{{\cal D}(l^*)-{\cal D}}$, 
indicating that $L(T)$ shown in Eq. (\ref{eq:lorenz_T}) is systematic. 
$L(T)$ for this case is shown in Fig.~\ref{fig:lorenz}. We emphasize that 
such a behavior is not expected in a higher-d FL case, in which the Lorenz 
number is constant even close to the MIT point\cite{castellani_lorenz}. 

{\it Lorenz number in dirty quantum wires.}
An experimental realization for the measurement of the transport 
properties of a LL is through a finite size wire connected to leads 
\cite{tarucha_quant_cond} or a carbon nanotube\cite{kim}. The theory 
presented here for a bulk material needs to be modified, as the boundary 
condition imposed by the contact is shown 
\cite{maslov_pure_wire,fazio_thermal_1d} to drastically influence the 
transport through the wire. Charge crosses the contact in terms of pieces 
of fractional charges \cite{safi}, whereas heat is carried by plasmon 
modes and they cross the contact like waves passing through barriers 
\cite{fazio_thermal_1d}. The Lorenz number will obtain a nontrivial wire 
size dependence as shown below. For simplicity we only discuss about 
spinless LL wires, and leave the discussion of the spinful case for a 
future publication \cite{lo}. Besides, we follow Ref. 
\cite{maslov_pure_wire} and model the leads as 1d noninteracting FLs. 

We first briefly recall the {\it clean} wire case. The electrical 
conductance $G$ is purely from the contact and is quantized 
\cite{maslov_pure_wire}, $G=G_0=e^2/2\pi$ (for $\hbar=1$), which implies 
a perfect transmission of the total charge after a long time (dc) 
measurement \cite{safi}. The mismatch of the plasmon wave velocities 
at the two sides of the contact leads to $T$-dependent thermal conductance 
${\cal K}_{\text{th}}$ normalized by $T$ \cite{fazio_thermal_1d}, which, 
at low temperatures $T\ll u/d$ ($d$ the wire size), asymptotically 
approaches to a constant ${\cal K}_{\text{th}}/T\simeq \pi/6$ 
\cite{fazio_thermal_1d}. The WF law is typically violated for a generic 
$T$ and is restored at $T\ll u/d$. 

Let us now turn to the case of a wire with  impurities. We consider the 
case of $l_{\rm loc}$ being much larger than $d$ and/or $l_{\text{th}}$ 
so that we can neglect Anderson localization. The total electrical 
resistance is the sum of the contact resistance $G_0^{-1}$ and the wire 
resistance $G_{\rm wire}^{-1}$, resulting in the following total 
conductance, 
\begin{eqnarray}
G=(G_0^{-1}+G_{\rm wire}^{-1})^{-1} \label{conductance_wire1}.
\end{eqnarray}
$G_{\rm wire}$ depends on the relative values of the two length scales 
$d$ and $l_{\text{th}}(T)$.
  
1) At $d\gg l_{\rm th}$ ($T\gg u/d$), a simple Ohm law holds for the 
electrical conductance, {\it i.e.}, $G_{\rm wire}=\sigma(T)/d=G_0 2\pi 
l_{\text{el}}/d$ with $\sigma(T)$ and $l_{\text{el}}$ being found in 
Eq.~(\ref{eq:sigma1-1}). Eq.~(\ref{conductance_wire1}) reduces, for 
$d\ll l_{\rm el}(T)$, to Maslov's perturbation result 
\cite{maslov_conductance1d}, $G\simeq G_0[1-d/(2\pi l_{\rm el})],$ 
and, for $d \gg l_{\rm el}(T)$, crosses to the Drude formula. In this 
regime, the thermal conductance of even a clean wire, 
${\cal K}_{\text{th}}$, from Eq.~(9) of Ref.~\cite{fazio_thermal_1d},
depends on the details of the barrier, so that the thermal conductance 
of the dirty wire becomes a rather non-universal function of disorder,
wire size, and temperature \cite{lo}.  

2) At $ d\ll l_{\text{th}}$ ($T\ll u/d$), Eq.~(9) of 
Ref.~\cite{fazio_thermal_1d} indicates that heat is carried by plasmon 
modes of wavelength much larger than the distance between the contacts, 
and thermal conductivity of the clean wire becomes universal. In this 
regime, we can make use of the $T\Leftrightarrow u/d$ equivalence 
\cite{ogata,maslov_conductance1d} to obtain an explicit $d$-dependent 
Lorenz number. According to Ogata and Fukuyama, the finite size effect 
can be taken into account within the memory function formalism by replacing 
$\omega$ in Eqs. (\ref{sigma}) and (\ref{kappa}) by $i2u/d$. This leads to 
\begin{eqnarray}
&& G(d) = {ue^2 \over \pi}\, {1\over 2u + 2u \,{\cal C}_c(u,K) \beta(d)},
\label{conductance-d} \\
&& {\cal K}_{\rm th}(d) = {\pi uT\over 3}\, {1\over 2u + 2u\, 
{\cal C}_{\rm th}(u,K) \beta(d)}, \label{thermalconductance-d}
\end{eqnarray}
with $\beta(d)=(d/l_{\text{loc}})^{3-2K}$, and ${\cal C}_c(u,K)$ and  
${\cal C}_c(u,K)$ being easily found by comparing 
Eqs.~(\ref{conductance-d}) and (\ref{thermalconductance-d}) with 
Eqs.~(\ref{eq:sigma1-1}) and (\ref{eq:kappa1-1}). In writing down 
Eq.~(\ref{conductance-d}) we have carefully taken into account the 
screening of the electric field coming from the leads \cite{kawabata}.
We recover a universal Lorenz number in this regime in the absence of 
impurities in agreement with \cite{fazio_thermal_1d}. In the presence of 
impurities, Eqs.~(\ref{conductance-d}) and (\ref{thermalconductance-d}) 
immediately yield 
\begin{equation}   \label{eq:lorenz1}
L(d)\simeq L_0 \, {1+{\cal C}_c(u,K) \, \beta(d) \over 
1+{\cal C}_{\text{th}}(u,K) \, \beta(d)}, 
\end{equation}
which is a function of the wire size since ${\cal C}_c(u,K) \ne 
{\cal C}_{\text{th}}(u,K)$ in the presence of interactions. Thus, 
deviation from WF law is obtained in finite size systems with impurities 
even for $T\ll u/d$. 

{\it Conclusion.} 
We have investigated the WF law in a disordered LL system. At high $T$ 
where  thermal effects cut off  Anderson localization, the Lorenz number 
is constant in $T$, and, in the spinless fermion case, smaller (larger) 
than the universal $L_0$ for attraction (repulsion) between the fermions. 
Its dependence on the LL exponents is different from the one obtained in 
a single strong impurity case \cite{kane_thermal_1d}. As $T$ goes down, 
Anderson localization effects induce an interplay between disorder and 
electronic interactions, which is responsible for the violation of the 
WF law. When a dirty 1d wire is connected to leads we find the Lorenz 
number becomes a function of the wire length.

{\it Acknowledgments.}
We are indebted to N.~Andrei, P.~J.~Hirschfeld, A.~Rosch, and 
Y.-J.~Wang for valuable discussion. We also thank T.~Giamarchi,
Y.~Suzumura, and I.~Vekhter for comments on the manuscript.


\begin{thebibliography}{10}

\bibitem{ziman_phonon_book}
J.~M. Ziman, {\em Electrons and Phonons} (Clarendon, Oxford, 1962);
A.~A. Abrikosov, {\em Fundamentals of the theory of metals} (Elsevier 
Science, Amsterdam, 1988).

\bibitem{castellani_lorenz}
C. Castellani {\it et~al.}, Phys. Rev. Lett. {\bf 59},  323  (1987).

\bibitem{hill}
R.~W. Hill {\it et~al.}, Nature {\bf 414}, 711 (2001).

\bibitem{emery_revue_1d}
V.~J. Emery,  in {\em Highly Conducting One-Dimensional Solids}, edited by
J.~T. Devreese, R.~P. Evrard, and V.~E. van Doren (Plenum Press, New York 
and London, 1979).

\bibitem{solyom_revue_1d}
J. S{\'o}lyom, Adv. Phys. {\bf 28}, 209 (1979).

\bibitem{vondelft_revue1d}
J. {von Delft} and H. Schoeller, Ann. Phys. (Leipzig) {\bf 7},  225  (1998).

\bibitem{apel_spinless}
W. Apel, J. Phys. C {\bf 15}, 1973 (1982).

\bibitem{apel_loc}
W. Apel and T.~M. Rice, J. Phys. C {\bf 16}, L271 (1982).

\bibitem{apel_impurity_1d}
W. Apel and T.~M. Rice, Phys. Rev. B {\bf 26}, 7063 (1982).

\bibitem{scha}
Y. Suzumura and H. Fukuyama, J. Phys. Soc. Jpn. {\bf 52}, 2870
(1983); {\sl ibid.} {\bf 53}, 3918 (1984).

\bibitem{giamarchi_loc_ref}
T. Giamarchi and H.~J. Schulz, Phys. Rev. B {\bf 37}, 325 (1988), 
and references therein.

\bibitem{spinchains}
A.~V. Sologubenko {\it et~al.}, Phys. Rev. B {\bf 62}, R6108
(2000); Phys. Rev. B {\bf 64}, 054412 (2001). 

\bibitem{spinpeierls}
Y. Ando {\it et~al.}, Phys. Rev. B {\bf 58}, R2913 (1998); J. Takeya 
{\it et~al.}, Phys. Rev. B {\bf 61}, 14700 (2000); {\sl ibid.}  {\bf 62},  
R9260  (2000); {\sl ibid.} {\bf 63},  214407  (2001).

\bibitem{ladder}
A.~V. Sologubenko {\it et~al.}, Phys. Rev. Lett. {\bf 84},  2714  (2000);
C. Hess {\it et~al.}, Phys. Rev. B {\bf 64},  184305  (2001);
K. Kudo {\it et~al.}, J. Phys. Soc. Jpn. {\bf 70},  437  (2001).

\bibitem{choi_thermal_tmtsf}
M.~Y. Choi, P.~M. Chaikin, and R.~L. Greene, Phys. Rev. B {\bf 34},  
7727 (1986).

\bibitem{luther_conductivite_disorder}
A. Luther and I. Peschel, Phys. Rev. Lett. {\bf 32},  992  (1974).

\bibitem{giamarchi_umklapp_1d}
T. Giamarchi, Phys. Rev. B {\bf 44},  2905  (1991).

\bibitem{rosch_transport1d}
A. Rosch and N. Andrei, Phys. Rev. Lett. {\bf 85},  1092  (2000).

\bibitem{goetze}
W. G{\"o}tze and P. W{\"o}lfle, Phys. Rev. B {\bf 6},  1226  (1972).

\bibitem{kane_thermal_1d}
C.~L. Kane and M.~P.~A. Fisher, Phys. Rev. Lett. {\bf 76},  3192  (1996).

\bibitem{giamarchi_impurities} T. Giamarchi and H. Maurey in {\em
Correlated fermions and transport in mesoscopic systems} edited by
T. Martin, G. Montambaux and J. Tran Thanh Van (\'Editions
Fronti\`eres, Gif sur Yvette, 1996).

\bibitem{fazio_thermal_1d}
R. Fazio, F.~W.~J. Hekking, and D.~E. Khmelnitskii, Phys. Rev. Lett. 
{\bf 80}, 5611  (1998).

\bibitem{krive_thermal1d}
I.~V. Krive, Fiz. Nizk. Temp. {\bf 24},  498  (1998), [Low Temp. Phys. 
{\bf 24}, 377 (1998)].

\bibitem{footnote2} Certain type of interactions in the spin channel 
may cause complication. This is left for a future publication.\cite{lo}

\bibitem{tarucha_quant_cond}
S. Tarucha, T. Honda, and T. Saku, Sol. State Comm. {\bf 94},  413  (1995).

\bibitem{kim}
P. Kim, L. Shi, A. Majumdar, and P. L. McEuen, Phys. Rev. Lett. {\bf 87}, 
215502 (2001). 

\bibitem{maslov_pure_wire}
D.~L. Maslov and M. Stone, Phys. Rev. B {\bf 52},  R5539  (1995).

\bibitem{safi} I. Safi and H. J. Schulz, Phys. Rev. B {\bf 52}, R17040 
(1995).

\bibitem{lo}
M.-R. Li and E. Orignac, in preparation (unpublished).

\bibitem{maslov_conductance1d}
D.~L. Maslov, Phys. Rev. B {\bf 52},  R14368  (1995).

\bibitem{ogata}
M. Ogata and H. Fukuyama, Phys. Rev. Lett. {\bf 73}, 468 (1994).

\bibitem{kawabata} 
A. Kawabata, J. Phys. Soc. Jpn. {\bf 65}, 30 (1996).

\end{thebibliography}
\end{document}